\documentstyle[aps]{revtex}
\begin{document}
\newcommand{\beq}{\begin{equation}}
\newcommand{\eeq}{\end{equation}}
\newcommand{\beqn}{\begin{eqnarray}}
\newcommand{\eeqn}{\end{eqnarray}}
\newcommand{\bmath}{\begin{mathletters}}
\newcommand{\emath}{\end{mathletters}}
\title{Superconductivity from Undressing}
\author{J. E. Hirsch }
\address{Department of Physics, University of California, San Diego\\
La Jolla, CA 92093-0319}
 
\date{\today} 
\maketitle 
\begin{abstract} 
Photoemission experiments in high $T_c$ cuprates indicate that 
quasiparticles are heavily 'dressed' in the normal state, 
particularly in the low doping regime. Furthermore these experiments show
that a gradual undressing occurs both in the normal state as the system 
is doped and the carrier concentration increases, as well as at
fixed carrier concentration as the temperature is lowered and the
system becomes superconducting. A similar picture can be inferred from
optical experiments. It is argued that these experiments
can be simply understood with the single assumption that the
quasiparticle dressing is a function of the local carrier concentration. 
Microscopic Hamiltonians describing this physics are discussed.
The undressing process manifests itself in both the one-particle
and two-particle Green's functions, hence leads to observable
consequences in photoemission and optical experiments respectively.
An essential consequence of this phenomenology is that the microscopic
Hamiltonians describing it break electron-hole symmetry: these
Hamiltonians predict
that superconductivity will only occur for carriers with hole-like
character, as proposed in the theory of hole superconductivity.

\end{abstract}
\pacs{}

\section{Introduction}
Photoemission experiments in high $T_c$ cuprates show that the spectral
function has two contributions: a coherent quasiparticle peak, and a 
broad incoherent background. Recently, Ding et al\cite{ding} have provided an insightful 
discussion of the phenomenology observed in photoemission in terms of the 
one-electron coherence factor, or quasiparticle weight, Z. The analysis
of Ding et al, as well as a variety of other analysis and experimental
data\cite{coh1,coh2,coh3}, suggest that the quasiparticle coherence increases in high $T_c$
cuprates both as the carrier concentration increases in the normal state,
as well as as the temperature decreases and the system becomes superconducting.

These observations can be simply understood by assuming that the quasiparticle
dressing in high $T_c$ cuprates is a $decreasing$ function of the 'local' carrier 
concentration,
more specifically of the probability of finding another carrier in the
vicinity of a given carrier. This local carrier concentration will increase
both as the total carrier concentration increases through doping, as well
as in a dilute carrier concentration regime when Cooper pairs form and
the superconducting state develops. This physics is qualitatively depicted in 
Figure 1.

We argue that the essential physics of the high $T_c$ phenomenon is
$undressing$\cite{color,polar1}. At low carrier density,
carriers are heavily dressed in the normal state, due to
coupling to a bosonic degree of freedom. When carriers pair and the
system becomes superconducting, carriers partially undress. Similarly,
when the system is doped in the normal state carriers increasingly
undress. This will occur if the coupling to the boson degree of freedom
is a function of the local carrier concentration, and becomes weaker as
the local carrier concentration increases. This feature, we propose,
 is what makes
the material a high temperature superconductor: carriers will
pair in order to undress, i.e. to reduce the coupling to this
boson degree of freedom. Paradoxically, by becoming $bound$ in Cooper
pairs, carriers become more $free$. At high carrier concentrations, 
carriers are undressed already in the normal state and hence superconductivity
does not occur.

The undressing will give rise to a lowering of the system's free
energy, and hence to the condensation energy of the superconductor.
Because it is an undressing transition, it is the $kinetic$  energy
that is lowered as the system becomes superconducting\cite{area}: 
as carriers undress,
their effective mass decreases, and this higher mobility in the
superconducting state is what provides the'glue' for the collective
order. Naturally, this will cause observable manifestations in optical
properties\cite{basov,fugol}.

There are two key distinct questions related to the understanding of
high $T_c$ superconductivity. One relates to the nature of the bosonic
degree of freedom coupled to the electrons that gives rise to pairing and other 
phenomena: is it magnetic, electronic, phononic, or other? What is its
spectral density? The second question
 is, what is the
essential physics of the phenomenon, and what are the key experimental signatures?
These two questions are not necessarily tightly coupled, and here we wish
to draw a sharp distinction between them. We do have definite views on the first
question, namely that the mechanism is purely electronic\cite{hole} rather than
magnetic, phonon, or a combination of various. However this paper mainly
focuses on the second question, for which we propose the undressing scenario.
It is possible that this  scenario
may be applicable independent of the detailed answer to the first question.

However, we argue also that the theory of hole 
superconductivity\cite{holesc1,holesc2} naturally leads to
the phenomenology described above, and, conversely also, that the
phenomenology described above naturally leads to the theory of
hole superconductivity.  The electron-hole symmetry
breaking leads to the undressing scenario, and, the assumption of
increased undressing with
increased carrier concentration leads to electron-hole symmetry breaking.
Furthermore, of the two possible choices, we argue that  it is $hole$ $carriers$
that are heavily dressed in the normal state, and that electron carriers instead
are lightly dressed, or undressed. As holes pair, or as the system is
doped with holes, holes become more 'like' electrons, and they undress.
We are not aware of any other microscopic
scenario that can describe the physics of 'undressing'. Through the
framework of the theory of hole superconductivity it is seen that 
another unavoidable consequence of this 'undressing' physics is
the prediction of tunneling asymmetry of universal sign\cite{tun}, as well
as of charge imbalance in the superconducting state\cite{imb}.

The microscopic physics is easily described qualitatively: there are carriers
in a band, and there is a background degree of freedom at each atom, or site.
Presence of a carrier  at a site will modify, or 'disrupt', the
background degree of freedom. So far, the physics is the same as
in a variety of electron-boson models. The essential distinguishing 
feature here is that the disruption caused by the first carrier on a site
is assumed to be $different$ to that caused by a second carrier on
that site. This essential feature leads to superconductivity
through undressing, with distinct characteristics.

We have proposed in the past a variety of microscopic Hamiltonians to
describe this physics: (1) a purely electronic model with two
orbitals per site\cite{twoorb}; (2) a model of electrons (one orbital per site)
coupled to a spin 1/2 degree of freedom at each site\cite{hole,holesc1}, and
(3) a variety of generalized Holstein models\cite{holst,pincus,polar2}. The essential physics
of all these Hamiltonians (in the parameter range considered) is the
same, qualitatively discussed above. We believe that either of these Hamiltonians
is a plausible choice for the description of the physics of electrons
in atoms with several electrons in the outer shell, which we believe
is the relevant physics: the 'background'
degree of freedom is simply a way of representing the excitations of
these other electrons. However, such Hamiltonians could also be used to
describe other microscopic physics, for example coupling to a high
frequency phonon.

Some of the consequences of the physics of undressing for optical 
properties have already been discussed\cite{color}. In particular, it led
to the prediction of apparent optical sum rule violation\cite{area}, which was
recently detected experimentally\cite{basov}. This is due to the fact that
undressing manifest itself in the two-particle Green's function.
However, as discussed in this paper, undressing also manifests
itself in the one-particle Green's function. Hence, 
it will have observable consequences for photoemission 
experiments. The close connection between the effects observed in
photoemission and in optical experiments has recently been emphasized
by Norman and coworkers\cite{norman}.

For most of this paper we will use a specific generalized Holstein
model\cite{pincus,polar2} to describe the physics of superconductivity through undressing.
The reason is simply that a large amount of theoretical work has been
done on the Holstein model. In the Holstein model usually considered\cite{holst}, the
electron-hole symmetric version, the physics of undressing is absent. 
Nevertheless, much of the mathematical treatment carries over to the
generalized model considered here.

Furthermore, we restrict ourselves to the small polaron regime of the model,
both because the physics in that regime is most transparent, and because
we believe it is the most appropiate regime to describe high $T_c$ cuprates.
Alexandrov and coworkers\cite{alex1,alex2} have performed a large amount of very 
interesting
theoretical work on the symmetric Holstein model in the small polaron
regime, and we will draw on some of their seminal contributions. In the
conclusion we will however discuss the fundamental difference in the
physics described by our model and that described by Alexandrov's work.

\section{The physics and Hamiltonians}
The physics is qualitatively depicted in Figure 2. A local bosonic degree of
freedom couples to the electron (or hole) at the site. The first electron
at the site causes a small distortion of the boson degree of freedom, the second 
causes a much larger distortion. Conversely, the first hole
at the site causes a large change of the boson degree of freedom, the second 
 a small one. The discussion can be consistently carried out with electrons
or with holes; we will use the language of holes in most of this paper
only because it is somewhat simpler to describe the physics of a
few holes rather than that of a lot of electrons. Notice that the key
physics of electron-hole asymmetry is introduced at the outset.

Let $|n>$ be the ground state of the local boson degree of freedom when there are n 
holes at the site, and $|n^l>$ the l-th excited state of that boson degree
of freedom, hence $|n> \equiv |n^0>$. Consider the ground state overlaps of the 
boson degree of freedom with different number of holes at the site:
\bmath
\beq
S=<0|1>
\eeq
\beq
T=<1|2>
\eeq
\emath
The physics of undressing arises if $S\neq T$\cite{color}. Specifically, for $holes$ to 
undress as the hole concentration increases,
\beq
S<T
\eeq
is required, that is, the overlap matrix element of the ground state of the boson
degree of freedom with zero and one hole is smaller than the one between one
hole and two holes. The effect will be strongest when $S<<T$; in an ideal
situation, one will have $T\sim 1$. The factors $S^2$ and $T^2$ are the 
quasiparticle weights of single hole carriers and single electron 
carriers respectively  in a Fermi liquid description: hole carriers will have
a small quasiparticle spectral weight, electron carriers a large quasiparticle
weight.

When a hole at site i is destroyed, with the boson degree of freedom in its
ground state, one of two things can occur: the boson degree of freedom may
make a transition to the ground state with one fewer hole, or end up in an
excited state. The first process is a diagonal transition, the second a 
non-diagonal one. If the site is singly occupied 
we have, with $c_{i\sigma}$ a hole destruction operator,
\beq
c_{i\uparrow} |\uparrow>|1>=|0>|1>=
|0>|0>S+\sum_{l\neq 0} |0>|0^l><0^l|1>
\eeq
so that S gives the probability amplitude for the diagonal transition. Here, the
site state is represented as a direct product of the hole occupation state
and the boson state. The first term in Eq. (3) conserves energy,
 preserves the phase of the
wave function and gives rise to a coherent process, associated with the quasiparticle
contribution to the spectral function; the second part gives the incoherent
contribution. Similarly, when a hole is destroyed at a site that is
occupied by two holes,
\beq
c_{i\uparrow } |\uparrow \downarrow>|2>=|\downarrow>|2>=
|\downarrow>|1>T+\sum_{l\neq 0} |\downarrow>|1^l><1^l|2>
\eeq
and the weight of the coherent process is given by T, which is assumed
to be larger
than S. Hence, coherence will increase as the number of doubly occupied
sites (by holes) increases, which will be the case both when hole doping and
when hole pairing occurs. The completeness relations
\beq
\sum_l|n^l><n^l|=1
\eeq
ensure that an increase in the weight of coherent processes is 
accompanied by a decrease in the weight of incoherent processes, and
the spectral function sum rule is satisfied.

Similarly, the amplitudes of the ground state to ground state transitions
determine the effective mass of the carriers. If t is the hopping
amplitude for a carrier in the absence of coupling to the boson degree of
freedom, the hopping amplitude for a single hole when there are no other
holes in either of the two sites involved in the hopping process is
\beq
t_2=tS^2,
\eeq
when there is one other hole it is
\beq
t_1=tST\equiv t_2+\Delta t
\eeq
and when there are two other holes it is
\beq
t_0=tT^2
\eeq
the latter being of course also the hopping amplitude for a single electron
when there are no other electrons at the two sites involved. The
hopping amplitudes are schematically shown in Fig. 3. Single hole carriers
are most heavily dressed and hence have a large effective mass $m_h^*$, related
to the hopping amplitude $t_2$ by
\beq
m_h^*=\frac{\hbar^2}{2t_2a^2}
\eeq
with $a$ the lattice spacing. 
When holes hop in the presence of other holes,
they do so with hopping amplitude $t_1>t_2$, as a partial undressing occurs.
The effective hopping Hamiltonian arising from
these transitions is
\bmath
\beq
H_{hop}=-\sum_{<i,j>,\sigma}t_{ij}^\sigma 
(c_{i\sigma}^\dagger c_{j\sigma}+h.c.)
\eeq
\beq
t_{ij}^\sigma=t[S^2+S(T-S)(n_{i,-\sigma}+n_{j,-\sigma})
+(T-S)^2 n_{i,-\sigma}n_{j,-\sigma}]
\eeq
\emath
The third term in Eq. (10b) will be negligible in the low hole concentration
regime for reasonable values of on-site repulsion. Hence the effective Hamiltonian,
including also an on-site and nearest neighbor Coulomb repulsion, is
\beq
H_{eff}=-\sum_{<i,j>,\sigma>}[t_2+\Delta t(n_{i,-\sigma}+n_{j,-\sigma})]
(c_{i\sigma}^\dagger c_{j\sigma}+h.c.)
+U\sum_i n_{i \uparrow}n_{i \downarrow}+V\sum_{<ij>}n_in_j
\eeq
The difference in hopping amplitudes
\beq
\Delta t = t_1-t_2=tS(T-S)
\eeq
gives rise to pairing and superconductivity\cite{holesc2}. The undressing and 
accompanying effective mass lowering when
pairing occurs gives rise to apparent violation of the optical sum rule as
discussed in detail elsewhere\cite{area,color}.

Furthermore, in the normal state the effective hopping for a hole will
be an increasing function of hole concentration $n$\cite{holesc2}, given by
\beq
t(n)=t_2+n\Delta t
\eeq
This will cause an expansion of the effective bandwidth as the hole
concentration increases, as shown schematically in Fig. 4. 
The conductivity sum rule yields for the integrated optical absorption
for intra-band
processes 
\beq
\int^{\omega_m} _0 d\omega \sigma_1(\omega) = \frac{\pi e^2n}{2m^*}
\eeq
with $\sigma_1$ the frequency-dependent real part of the conductivity
 (per site),
$n$ the number of carriers per site, $m^*$ the effective mass and  $\omega_m$ 
a high frequency cutoff that allows only for intra-band processes. 
 Using Eqs. (9) and (13), 
\beq
\int^{\omega_m} _0 d\omega \sigma_1(\omega) =
\frac{\pi e^2 a^2 t(n) n}{\hbar ^2}
\eeq
so that the integrated low frequency spectral weight
increases faster than linearly with carrier concentration. 
There is evidence
from optical experiments that the low frequency optical spectral weight increases
with doping more rapidly than expected from the added number of carriers\cite{uchida},
in support of Eq. (15). This transfer of optical spectral weight from high
to low frequencies with hole doping is a manifestation of the undressing
that occurs with increasing hole concentration, and will be accompanied by
 a decrease in the spectral weight
of nondiagonal transitions, i.e.  hopping processes where the background 
degrees of freedom end up in excited states rather than the ground state.

The relation between quasiparticle spectral weight and effective mass
discussed above follows of course from general properties of 
many-body systems. The exact Green's function for a many-body system
can be written as
\beq
G(k,\omega)=\frac{1}{\omega-\epsilon_k-\Sigma(k,\omega)}
\eeq
where $\Sigma$ is the self-energy, and $\epsilon_k$ is measured from the
chemical potential. In the models considered here the self-energy has
no $k$-dependence, and we have for its real part 
\beq
\Sigma_{re}(\omega)=\Sigma_{re}(0)+\omega\frac{\partial \Sigma_{re}}
{\partial \omega}
\eeq
$\Sigma_{re}(0)$ just renormalizes the chemical potential. Hence
\beq
G(k,\omega)=\frac{1}{\omega(1-\frac{\partial \Sigma_{re}}{\partial \omega})
-\epsilon_k}+G'=\frac{Z}{\omega-Z\epsilon_k}+G'
\eeq
with
\beq
Z=\frac{1}{1-\frac{\partial \Sigma_{re}} {\partial \omega}}
\eeq
The term $G'$ contains the imaginary part of the self-energy and gives
rise to the incoherent contribution. Eq. (18) shows that the same
factor $Z$, the wave function renormalization factor
 that results from the frequency dependence of the
real part of the self-energy, determines the quasiparticle spectral weight and the
effective mass renormalization. In the models considered here, $Z$ is
a function of carrier concentration and we have simply
\beq
Z(n)=S^2+nS(T-S)+\frac{n^2}{4}(T-S)^2
\eeq
in the normal state, with $n$ the hole concentration per site.
In particular $Z=S^2$ or $Z=T^2$ for an almost filled
band ($n\rightarrow 0$) or an almost empty band ($n\rightarrow 2$)
respectively. In the superconducting state, $Z$ will increase as the
pair amplitude develops.

One Hamiltonian that describes this physics is the spin-fermion
model\cite{hole,holesc1} with site Hamiltonian given by\beq
H_i=(V(n_{i\uparrow}+n_{i\downarrow})-\omega_0)\sigma_z^i+
\Delta \sigma_x^i + Un_{i\uparrow}n_{i\downarrow}
\eeq
with $\sigma^i$ a spin 1/2 degree of freedom, and $n_{i\sigma}$ a hole
occupation operator. The physics described here arises in the
parameter regime $V>\omega_0$, $\Delta<<\omega_0$. Then, the spin
degree of freedom points approximately up when there is no hole
at the site, and approximately down when there are one or two holes
at the site, as shown schematically in Fig. 3. The change in the spin
state is much larger between hole occupation 0 and 1 than between
hole occupation 1 and 2, hence $S<<T$.

Another Hamiltonian that describes this physics is an electronic
model with two orbitals per site\cite{twoorb}, and a site Hamiltonian
\beq
H_i=Un_{i\uparrow}n_{i\downarrow} +U'n'_{i\uparrow}n'_{i\downarrow}
+Vn_{i}n'_{i} +\epsilon n'_i -t'(c_{i\sigma}^\dagger c'_{i\sigma}+h.c.)
\eeq
where the primed and unprimed operators refer to $electrons$ in the two site orbitals. 
The physics of undressing will arise if the condition
\beq
U'+2\epsilon < V-\epsilon <U
\eeq
is satisfied, together with the ordering $U,U',V>>\epsilon>>t'$.
These conditions ensure that a single electron resides primarily in the
lower level (unprimed orbital) while two electrons reside dominantly
in the higher level (primed orbitals). When an electron leaves a doubly
occupied orbital, the second electron relaxes to the lower energy level
giving a large renormalization effect, and $S<<1$, while $T=1$. This Hamiltonian 
in the parameter regime described may be justified from first principles
atomic physics calculations\cite{diat}.

Finally, the physics described here will arise in a variety of generalizations
of the Holstein Hamiltonian\cite{polar2} describing electrons interacting with a
local displacement degree of freedom $q_i$
\beq
H_i=\frac{p_i}{2m(n_i)}+\frac{1}{2}K(n_i)q_i^2+
\alpha(n_i)q_in_i +Un_{i\uparrow}n_{i\downarrow} .
\eeq
The undressing effect arises when the usual Holstein model is
generalized by allowing for an occupation dependence of the 
parameters, m, K or $\alpha$ in Eq. (24), or by adding an
anharmonic term $\beta q^4$ to the site Hamiltonian.
 In the remainder of this paper we will use one
particular version of this model, where the coupling $\alpha$ depends on
site occupation.

\section{A generalized Holstein model}

We consider the site Hamiltonian
\beq
H_i=\frac{p^2}{2m} + \frac{1}{2} K q^2 +
q[\alpha(n_\uparrow+n_\downarrow)-
\alpha ' n_{\uparrow}n_{\downarrow}]+Un_{\uparrow}n_{\downarrow}
\eeq
where the site index on the right-hand side is understood. The new term in
this Hamiltonian, proportional to $\alpha '$, breaks electron-hole
symmetry and gives rise to the physics of undressing. This term may be understood
as arising from a dependence of the electron-boson coupling on the
hole occupation ($\alpha(n_i)=\alpha-\alpha '(n_\uparrow+n_\downarrow-1)/2$).
 Assuming $\alpha, \alpha ' >0$, Eq. (25) implies that
the electron-boson coupling becomes weaker as holes are added. 
Alternatively, the new term may be understood as a modification of the on-site
Coulomb repulsion by the boson displacement, $U(q)=U-\alpha 'q$.
In terms of boson creation and annihilation operators the Hamiltonian is
\beq
H=\hbar \omega (a^\dagger a +\frac{1}{2})+M(a^\dagger +a)
[n_\uparrow+n_\downarrow-
\frac{\alpha '}{\alpha} n_{\uparrow}n_{\downarrow}]+Un_{\uparrow}n_{\downarrow}
\eeq
with $M=\alpha (\hbar \omega_0/2K)^{1/2}$, $\omega _0=(K/m)^{1/2}$. 
Our Hamiltonian differs
from the usual Frohlich-type Hamiltonian in that the boson couples to a term that is
quartic in fermion operators in addition to the usual quadratic one.\cite{sham}

By
completing the squares in Eq. (25), we obtain the effective 
on-site repulsion
\beq
U_{eff}=U-\frac{2\alpha^2+\alpha'^2-4\alpha \alpha'}{2K}
\eeq
We assume $\alpha' \leq \alpha $. Note that $\alpha '$ causes an $increase$
in the on-site repulsion; in particular, for $\alpha '>(2-\sqrt{2})\alpha$, the reduction
in $U$ due to the electron-boson coupling $\alpha$ is completely offset,
and $U_{eff}>U$. Still, the model in that regime will give rise to
superconductivity due to the undressing effect produced by $\alpha '$.
The polaron site energy is given by
\beq
\epsilon_0=-\frac{\alpha^2}{K}
\eeq

The equilibrium position of the oscillators with $n$ holes at the sites,
$q_n$, is
\bmath\beq
q_0=0
\eeq
\beq
q_1=-\frac{\alpha}{K}
\eeq
\beq
q_2=-\frac{2\alpha}{K}+\frac{\alpha '}{K}
\eeq
\emath
and the ground state overlap matrix elements for the states with $n$
and $n'$ holes is
\beq
<n|n'>=e^{-\frac{K}{4\hbar \omega_0}(q_n-q_{n'})^2}
\eeq
so that 
\bmath
\beq
S=e^{-\frac{\alpha ^2}{4K\hbar \omega_0}}\equiv e^{-\frac{g^2}{2}}
\eeq
\beq
T=e^{-\frac{(\alpha -\alpha ')^2}{4K\hbar \omega_0}}
\equiv e^{-\frac{g^2}{2}(1-\gamma)^2}
\eeq
\emath
with
\bmath
\beq
g=\frac{\alpha}{\sqrt{2K\hbar \omega_0}}=\frac{M}{\hbar \omega_0}
\eeq
\beq
\gamma=\frac{\alpha '}{\alpha}
\eeq
\emath
Hence, $T>S$, and in particular $T=1$ for $\alpha'=\alpha$. For that
particular case, $U_{eff}=U+\alpha^2/2K$. The effective low energy
Hamiltonian to first order in the bare hopping t is
\beq
H_{eff}=-\sum_{(i,j)}(t_2+\Delta t(n_{i,-\sigma}+n_{j,-\sigma}))
(c_{i\sigma}^\dagger c_{j\sigma}+h.c.)+U_{eff}\sum_i
n_{i \uparrow}n_{i \downarrow}
\eeq
with $t_2$ and $\Delta t$ given by Eqs (6) and (12).
Superconductivity will occur if the condition\cite{holesc2}
\beq
\frac{T}{S}>(1+\frac{U_{eff}}{D_h})^{1/2}
\eeq
is satisfied, with $D_h=2zt_2$ the single hole bandwidth
(z is the number of nearest neighbors to a site). Hence for 
$T\sim 1$ and sufficiently small $S$, superconductivity will occur for 
arbitrarily large values of the on-site repulsion $U_{eff}$.
Some discussion of the effect of a nearest neighbor Coulomb 
repulsion and of higher order corrections to the parameters in
$H_{eff}$ is given in ref. 18.
Matrix elements between ground state oscillator states and
excited states with different occupation number are given by
\bmath
\beq
<0|1^l>=<1|0^l>=\frac{e^{-g^2/2}g^l}{(l!)^{1/2}}
\eeq
\beq
<1|2^l>=<2|1^l>=\frac{e^{-g^2(1-\gamma)^2/2}(g(1-\gamma))^l}{(l!)^{1/2}}
\eeq
\emath

To calculate the single particle Green's function and spectral function
it is useful to formulate the problem in terms of polaron operators,
as is done in the usual Holstein model. In the absence of hopping
between sites, the Hamiltonian is diagonalized by a generalized
Lang-Firsov transformation\cite{lang,polar1}
\beq
\bar{H}_i=e^GH_ie^{-G}
\eeq
with 
\beq
G=g(a^\dagger - a)(n_\uparrow +n_\downarrow -\gamma
n_{\uparrow}n_{ \downarrow})
\eeq
with $a^\dagger$ a boson creation operator
\beq
a^\dagger=\frac{1}{\sqrt{2}}(\sqrt{\frac{m \omega_0}{\hbar}}q+
i\frac{1}{\sqrt{m\hbar \omega_0}}p)
\eeq
Boson operators transform according to 
\beq
\bar{a}=e^Gae^{-G}=a-g
(n_\uparrow +n_\downarrow -\gamma
n_{\uparrow}n_{ \downarrow})
\eeq
and fermion operators transform according to
\bmath
\beq
\bar{c}_{i \sigma}=e^Gc_{i\sigma}e^{-G}=c_{i\sigma} X_{i\sigma}
\eeq
\beq
X_{i\sigma}=e^{-g(a^\dagger-a)(1-\gamma n_{i, -\sigma})}
\eeq
\emath
In contrast to the usual Lang-Firsov transformation, the "dressing" operators
$X_{i\sigma}$ here depend of fermion occupation number. The hopping part
of the Hamiltonian
\beq
H_{hop}=-t\sum_{(i,j)} (c_{i\sigma}^\dagger c_{j\sigma}+h.c.)
\eeq
becomes
\beq
\bar{H}_{hop}=-t\sum_{(i,j)} (X_{i\sigma}^\dagger X_{j\sigma}
c_{i\sigma}^\dagger c_{j\sigma}+h.c.)
\eeq
The expectation value of the X-operators in the zero boson subspace is
\beq
<X_{i\sigma}>=e^{-\frac{g^2}{2}(1-\gamma n_{i,-\sigma})^2}
\eeq
i.e. the overlap matrix elements $S$ or $T$ of Eq. (31) depending on
the value of the hole occupation $n_{i,-\sigma}$.

\section{Exact results}
Consider a system with a single site. The single particle spectral function
at zero temperature for holes with spin up is given by
\beq
A^\uparrow _{n_\uparrow n_\downarrow}(\omega)
=2\pi\sum_l[|<l|c_\uparrow^\dagger|0>|^2
\delta(w-(E_l^{(n+1)}-E_0^{n}))+
|<l|c_\uparrow|0>|^2 \delta(w+(E_l^{(n-1)}-E_0^{n}))]
\eeq
where $|0>$ is the ground state of the system with
$n_\sigma $ holes of spin $\sigma$, and $|l>$ are states of the system with 
$n_\uparrow +1$ or $n_\uparrow-1$ holes with spin up. Using Eq. (3) we obtain
\beq
A^\uparrow _{n_\uparrow 0}(\omega)=2\pi S^2\delta(\omega -\epsilon_0)+2\pi\sum_{l\neq 0}
|<1^l|0>|^2[(1-n_\uparrow )\delta(w-\epsilon_0 -\omega_l)+
n_\uparrow \delta(w-\epsilon_0 +\omega_l)]
\eeq
with $\epsilon_0$ the site energy for one hole
and $\omega_l$ the energy of the l-th excited state of the boson. In a many body system
the spectral function has the form\cite{mahan}
\beq
A(k,\omega)=2\pi Z(k) \delta(\omega-(E_k-\mu))+A_{inc}(k,\omega)
\eeq
that is, a sharp $\delta$-function describing the quasiparticle and
an incoherent background $A_{inc}$. The weight of the $\delta$-function, 
$0\leq Z\leq 1$, gives the degree of coherence. For our case Eq. (45),
the first term, corresponding to diagonal transitions, represents the
coherent part, and the second term is the incoherent part, hence $Z=S^2$.
If instead the ground state of the site has a spin down hole the
spectral function is
\beqn
A^\uparrow _{n_\uparrow 1}&=&2\pi T^2\delta(\omega -\epsilon_0-U_{eff})
+
2\pi\sum_{l\neq 0} [|<2^l|1>|^2
[(1-n_\uparrow )\delta(w-\epsilon_0-U_{eff} -\omega_l)\nonumber \\
&+&
n_\uparrow \delta(w-\epsilon_0-U_{eff} +\omega_l)]
\eeqn
where $(2\epsilon_0+U_{eff})$ is the energy of two holes at the site.
Here, the quasiparticle weight is $Z=T^2$. We assumed the
operators in Eq. (44) to be hole operators, and $S<T$. Then, as holes
are added to the site the coherent part of the spectral function
increases and the incoherent part decreases, so that the sum rule
\beq
\int d \omega A_n(\omega)=2\pi
\eeq
is satisfied.

For the case of the generalized Holstein model defined in the previous
section, $\omega_l=l\omega_0$ and
\bmath
\beq
A^\uparrow _{n_\uparrow 0}(\omega)=2\pi e^{-g^2}\delta(\omega -\epsilon_0)+
2\pi e^{-g^2}\sum_{l=1}^\infty
\frac{g^{2l}}{l!}[(1-n_\uparrow )\delta(w-\epsilon_0 -l\omega_0)+
n_\uparrow \delta(w-\epsilon_0 +l\omega_0)]
\eeq
\beqn
A^\uparrow _{n_\uparrow 1}(\omega)&=&2\pi e^{-g^2(1-\gamma)^2}\delta(\omega -\epsilon_0-U_{eff})
+2\pi e^{-g^2 (1-\gamma)^2} \times \nonumber \\
& &\sum _{l\neq 0} 
[\frac{g^{2l}(1-\gamma)^l}{l!}
[(1-n_\uparrow )\delta(w-\epsilon_0-U_{eff} -l\omega_0)+
n_\uparrow \delta(w-\epsilon_0-U_{eff} +l\omega_0)]
\eeqn
\emath
The qualitative behavior is shown in Fig. 5. Similar qualitative behavior
is found for the two other models discussed in Sect. II.

This simple example bears directly on the understanding of
qualitative features of 
photoemission experiments in high $T_c$ cuprates. When the system has
a low concentration of holes in the normal state, the spectral
function will look qualitatively like Fig. (5a).  The $l=0$ peak,
which gives the quasiparticle contribution, will be nearly absent if
$g^2$ is large. The other peaks in an extended system will merge into
a continuum  incoherent contribution, peaking at some high energy.
Both when the system is doped with holes, and at fixed doping when
the temperature is lowered and superconductivity sets in, the number
of doubly occupied (by holes) sites will increase, and hence the
spectral function will have a larger contribution of 
 Fig. 5(b). Hence the intensity of the quasiparticle peak in 
the spectral function will increase both as the system goes
superconducting and as the system is doped in the normal state.

Similarly, consider the optical absorption resulting from transitions
between localized particles at sites 1 and 2. Assume the current operator
is given by\cite{mahan}
\beq
j=\sum_\sigma p_{12} c_{2\sigma}^\dagger c_{1\sigma} +h.c.
\eeq
The real part of the optical conductivity at zero temperature is given by
\beq
\sigma_1(\omega)=\frac{\pi}{\omega}\sum_m|<m|j|0>|^2\delta(\omega-(E_m-E_0))
\eeq
For the case where there is one hole at site 1 and no hole at site 2 in the
ground state, Eq. (51) yields
\beq
\sigma_1^{(1)}(\omega)=
\frac{\pi}{\omega}p_{12}^2 [ S^4\delta(\omega)+
\sum_{(l l')\neq (00)}|<0^ll1>|^2|<1^{l'}|0>|^2\delta(\omega-
\omega_l-\omega_{l'})]
\eeq
and for the case where there is one hole at each site, of opposite spin
\beq
\sigma_1^{(2)}(\omega)=
\frac{2\pi}{\omega}p_{12}^2[ S^2T^2\delta(\omega-U_{eff})+
\sum_{(l l')\neq (00)}|<0^l|1>|^2|<2^{l'}|1>|^2\delta(\omega-
U_{eff}-\omega_l-\omega_{l'})]
\eeq
Once again we have separated the coherent contribution, involving no
excited bosons, from the incoherent contribution where bosons are emitted. For the
generalized Holstein model, using the relation
\beq
\sum_{l=0}^L\sum_{l'=0}^{L-l} \frac{g_1^{2l}g_2^{2l'}}{l!l'!}=
\frac{(g_1+g_2)^L}{L!}
\eeq
these relations become
\bmath
\beq
\sigma_1^{(1)}(\omega)=\frac{\pi}{\omega}p_{12}^2
[e^{-2g^2}\delta(\omega)+e^{-2g^2}\sum_{l=1}^\infty\frac{(2g^2)^l}
{l!} \delta(\omega-\omega_0l)]
\eeq
\beq
\sigma_1^{(2)}(\omega)=\frac{2\pi}{\omega}p_{12}^2
[e^{-g^2[1+(1-\gamma)^2]}\delta(\omega-U_{eff})+
e^{-g^2[1+(1-\gamma)^2]}\sum_{l=1}^\infty
\frac{g^{2l}[1+(1-\gamma)^2]^l}
{l!} \delta(\omega-U_{eff}-\omega_0l )]
\eeq
\emath
The qualitative behavior of these quantities is similar to that of the
single particle spectral functions shown in Fig. 5. In an extended
system the zero boson terms give rise to the intra-band optical absorption.
According to these results, for $\gamma \neq 0$ the intra-band absorption
will increase as holes are added to the system more rapidly than proportional to
the number of holes, since there is an increasing contribution of
$\sigma_1^{(2)}(\omega)$ relative to $\sigma_1^{(1)}(\omega)$ . Correspondingly,
the contribution of $l\neq 0$ terms, corresponding to non-intra-band
processes, decreases. The two other models discussed in section II display
similar physics\cite{color}.

Furthermore, as the system with a dilute concentration of holes becomes
superconducting, the relative contribution of $\sigma_1^{(2)}$ will also 
increase relative to that of $\sigma_1^{(1)}$, because there is an increased
fraction of configurations with holes on the same or on nearest neighbor
sites. In that case the resulting extra
intra-band optical spectral weight goes into the $\delta$-function that
determines the London penetration depth, and an apparent violation of
the Ferrell-Glover-Tinkham conductivity sum rule results.\cite{area,color}

In summary, in the two simple examples discussed in this section it is seen
that the process of undressing will manifest itself similarly in the
one-particle and two-particle spectral functions, which determine the
results of photoemission and optical absorption measurements. For
both types of observables, within the class of models discussed here,
an increase in the local hole concentration through doping or through
pairing gives rise to increased coherence.

\section{Spectral function in the dilute limit}
To obtain further insight on the effect of undressing in connection with
superconductivity we consider the dilute limit, that is, the limit when
the number of hole carriers in the band goes to zero. The wavefunction
for a single pair of holes governed by the effective Hamiltonian Eq. (11) 
can be found exactly and is of the form\cite{dilute}
\bmath
\beq
|\Psi>=\frac{1}{\sqrt{N}}\sum_k f_k c_{k\uparrow}^\dagger
c_{-k\downarrow}^\dagger |0>=\frac{1}{\sqrt{N}}\sum_{i,j}f(i-j)
c_{i\uparrow}^\dagger c_{j\downarrow}^\dagger|0>
\eeq
\beq
f_k=\sum_je^{iR_j}f(j)
\eeq
with
\beq
\sum_\delta |f(\delta)|^2=1
\eeq
\emath
Equations that determine the amplitudes $f(\delta)$ and the pair
binding energy $E_b$ are found from exact solution of the
Schrodinger equation for a single pair\cite{dilute}. In particular,
in the limit where $t_2\rightarrow 0$ in Eq. (11) only the amplitudes
$f(\delta)$ for on-site and nearest neighbor pairs are nonvanishing\cite{strong}.

We consider the part of the spectral function Eq. (44) corresponding 
to the destruction of a hole of momentum k. Applying the operator
\beq
c_{k\uparrow}=\frac{1}{\sqrt{N}} \sum_k e^{ikR_i} c_{i \uparrow}
\eeq
to the wavefunction Eq. (56a) gives rise to diagonal and
non-diagonal terms, as given by Eqs. (3) and (4). The diagonal terms, where
the bosons remain in the ground state, yields
\beq
(c_{k\uparrow}|\Psi>)_{diag}=\frac{1}{\sqrt{N}}
[(T-S)f(0)+Sf_k]c_{-k \downarrow}|0>
\eeq
and the non-diagonal terms are
\beqn
(c_{k\uparrow}|\Psi>)_{non-diag}&=&\frac{1}{N}\sum_{i,l}e^{ikR_i}
f(0)|\downarrow>_i<1^l|2>\nonumber \\
&+& \frac{1}{N}\sum_{i\neq j,l}e^{ikR_i} f(i-j)|\downarrow>_j|0^l>_i<0^l|1>
\eeqn
Assuming that states with excited bosons in a real space representation
are approximately eigenstates of the Hamiltonian one obtains the
spectral function
\beqn
A_s(k,\omega)&=&2\pi n_p[[(T-S)f(0)+Sf_k]^2\delta(\omega +\epsilon_k+E_b)\nonumber \\
&+& \sum_{l\neq 0}[ |<1^l2>|^2f^2(0)+|<0^l|1>|^2 (1-f(0)^2)]\delta (\omega
+\epsilon_k+E_b+\omega_l)]
\eeqn
We have assumed that there are $n_p$ pairs in the ground state that
don't interfere with one another. In contrast, in the standard models with
no electron-hole symmetry breaking such as the conventional
Holstein model, $S=T$ and Eq. (60) becomes
\beq
A_s(k,\omega)=2\pi n_p[S^2f_k^2\delta(\omega +E_b)+
\sum_{l\neq 0} |<0^l|1>|^2 \delta (\omega
+\epsilon_k+E_b+\omega_l)]
\eeq

Comparing Eqs. (60) and (61) shows clearly the effect of
electron-hole symmetry breaking. The coherent part of the
spectral function in the superconducting state will be
enhanced when $T>S$ proportionally to the on-site
pair wave function amplitude $f(0)$. As a consequence the
enhancement will be larger the smaller the coherence length, i.e.
the size of the pair wave function. Correspondingly, the
incoherent part of the spectral function is reduced, since
$|<1^l|2>| < |<0^l|1>| $ if $S<T$, as discussed in Sect. II.
Furthermore, the enhancement is proportional to the number
of pairs. The number of pairs within a two-fluid model is
given by the superfluid density, $n_p=\lambda_L(0)^2/\lambda_l(T)^2$,
with $\lambda_L$ the London penetration depth, and grows
proportionally to the gap squared as the temperature is lowered below
$T_c$. Also, the superfluid density will increase with doping.
Hence in the presence of the undressing effect we conclude that there
will be an extra contribution to the coherent part of the spectral
function that increases both as the temperature is lowered and
superconductivity sets in, and as the doping is increased in the 
superconducting state. These
effects should be observable in photoemission experiments. We discuss
these effects in more detail after considering the spectral function
in the normal state in the next section.

\section{Spectral function in the normal state}

The calculation of the spectral function in the normal state of the
generalized Holstein model  follows closely the 
treatment of Alexandrov and Ranninger for the electron-hole
symmetric case\cite{alex2}. In the single particle Green's function
\beq
G(m,\tau)=-<T_\tau c_{0\uparrow}(\tau)c_{m\uparrow}^\dagger (0)>
\eeq
fermion operators are replaced by the transformed operators Eq. (35),
and averages over fermions and bosons are decoupled. The result for the
spectral function is
\beqn
A(k,\omega)&=&2\pi e^{-g^2}[[1+n(e^{\gamma g^2}-1)]
\delta(\omega + (\epsilon_k-\mu))\nonumber \\
&+&
\frac{1}{N}\sum_{k'}\sum_{l=1}^\infty \frac{g^{2l}}{l!}
[1+n[e^{\gamma g^2}(1-\gamma)^l-1]]\times \nonumber \\
&[&(1-n_{k'})\delta (\omega+l\omega_0+\epsilon_{k'}-\mu)+
n_{k'}\delta(\omega-l\omega_0+\epsilon_{k'}-\mu)]
\eeqn
where $n$ is the hole concentration.
For $\gamma=0$ it reduces to the result of Alexandrov and Ranninger.
The first term describes the coherent contribution, the second term
the incoherent one. When $\gamma >0$, the coherent part increases
at the expense of the incoherent part as the hole concentration
$n$ increases. The amplitude of the coherent part of the spectral function
is proportional to the effective bandwidth, which is given (to lowest order in 
$\gamma$) by
\beq
D(n)=2zt(n)=2z(t_2+n\Delta t)=2zte^{-g^2}(1+n(e^{\gamma g^2}-1))
\eeq
and increases linearly with doping.

In a photoemission experiment
the quantity probed is
\beq
I_0(k,\omega)=A(k,\omega)f(\omega)
\eeq
with $f(\omega)$ the Fermi function. To take into account experimental
resolution we calculate the convolution
\beq
I(k,\omega)=\int d\omega ' F(\omega -\omega ') I_0(k,\omega ')
\eeq
with $F(\omega)$ a Gaussian function with width $\sigma _\omega$. 
An example of the behavior obtained is shown in Fig. 6.
Note that the quasiparticle peak is absent for low hole
concentration and appears as the hole concentration increases.

\section{Quasiparticle spectral weight}

Here we summarize the qualitative conclusions on the quasiparticle
spectral weight that we can draw from the results of the previous 
two sections. In the normal state, the quasiparticle weight Z
is, from Eqs. (63) and (46)
\beq
Z_n=S^2(1+n(\frac{T}{S}-1))=e^{-g^2}(1+n(e^{\gamma g^2}-1))
\eeq
The first form is general for the class of models discussed in
 section II, the second the particular case of the generalized Holstein model.
In the superconducting state the quasiparticle weight, from Eq. (60), is
\beq
Z_s=n_p [(T-S)f(0)+Sf_k]^2
\eeq
For the particular case $T>>S$ and the generalized Holstein model,
\beq
Z_s=n_p e^{-g^2}(e^{\gamma g^2}-1)^2 f^2(0)
\eeq
Equations (67) and (69) show that the quasiparticle weight will
grow both as the system is doped in the normal state and as the
system goes into the superconducting state. However, because the
factor $(e^{\gamma g^2}-1)$ appears $squared$ in the expression for $Z_s$,
the onset of coherence will be more pronounced as the temperature
 is lowered than as
$n$ is increased in the normal state if that factor is large,
which will be the case for $S<<1$ and $T>>S$. This appears to be the
observation in photoemission experiments\cite{ding}. More generally,
the conditions $S<<1$ and $T>>S$ are consistent with the observations of a
high degree of incoherence in the underdoped regime of cuprates and onset of coherence
for a relatively modest increase in $n$. Furthermore, in the model of
hole superconductivity that condition, which is equivalent to 
$\Delta t>>t_2$, and $t_2$ very small, is required to fit a variety of
experiments\cite{holesc2}.

In that case, one may expect to see a pronounced quasiparticle peak 
emerge as superconductivity sets in even in the underdoped regime.
The temperature dependence of $Z_s$ is governed by $n_p$, the superfluid
weight, which is what is observed experimentally\cite{ding}.
The amplitude of the pair wave function $f(0)$ is found to be essentially
independent of temperature in the model of hole superconductivity\cite{holesc2}.

Furthermore, the doping dependence of the quasiparticle peak in the superconducting
state Eq. (69) will be determined by that of the superfluid weight $n_p$, and also by
the doping dependence of $f(0)$. In the model
of hole superconductivity the superfluid weight is found to grow approximately
linearly with doping\cite{area}. The on-site pair amplitude $f(0)$ is found
to decrease as the system moves towards the overdoped regime because there 
is a cross-over to the weak coupling regime and the coherence length
increases\cite{holesc2,strongweak}. Ding et al\cite{ding} find Z in the
superconducting state to increase linearly with doping in the underdoped
regime, and taper off on the overdoped side. Hence this behavior is
qualitatively consistent with the prediction of Eq. (69). Detailed
quantitative comparison will be given elsewhere.

Finally, the expression Eq. (67) for the quasiparticle weight in the normal
state predicts that a quasiparticle peak should be seen in the normal state
for sufficiently high doping, which is also consistent with observations\cite{ding}.

\section{Conclusions}
We have considered in this paper a class of models that appear to 
contain  key ingredients of the
phenomenology inferred from a wide range of photoemission
and optical experiments in high $T_c$ cuprates. This phenomenology is that increased 
coherence is
observed both in the normal state when the 
hole concentration increases, and for fixed hole concentration when the
temperature is reduced and the system goes superconducting. The emergence
of coherence thus appears to be associated with an increase in the
'local' carrier concentration. We have pointed out that this 
phenomenology follows from the class of models considered in the
theory of hole superconductivity. In these models, the first hole on a site
causes a large disruption in a background degree of freedom, the second
hole a much smaller disruption. Electron-hole asymmetry thus results
as an essential feature. The resulting  low energy
effective Hamiltonian contains the term $\Delta t$, which leads to
superconductivity and is a manifestation of the undressing process.
We illustrated this physics with some simple examples.

An explicit calculation of the single particle
spectral function for one such model, a generalized
Holstein model, showed that indeed it exhibits a decrease in
the incoherent contribution and the emergence of a quasiparticle peak
in the normal state as the hole concentration increases. Furthermore, we
showed in the particular case of the dilute limit that
increased coherence also results in the superconducting state
as a result of the formation of the pair wave function.
An  approximate calculation of the one-particle spectral weight 
in the superconducting state  for a more general case
will be discussed in a forthcoming paper.

These models have similar consequences for the two-particle Green's function,
and predict that
a transfer of optical spectral weight from high to low frequency in the
normal state should occur as the
hole doping increases due to the undressing process. 
Furthermore,  in the superconducting state 
the undressing process gives rise to  an 'apparent'
violation of the Ferrell-Glover-Tinkham low frequency optical sum rule 
\cite{area,basov} accompanied by a decrease in the optical
absorption at high frequencies\cite{color,fugol}, signaling the fact
that the superconducting condensation energy is $kinetic$ energy.  

Other models have been recently discussed in the literature where
the superconducting condensation energy is argued to come from
kinetic energy\cite{ander,kivel}. However none of those models address 
the questions of from where, and through which mechanism, the transfer of
high frequency optical spectral weight occurs.

Even though the Hamiltonian considered here is related to the model
studied by Alexandrov in his theory of bipolaronic 
superconductivity\cite{alex1},
there are fundamental differences. Alexandrov's model, by not
breaking electron-hole symmetry, does not contain the physics of
undressing. Also, his model predicts pair formation (bipolarons)
in the normal state, which subsequently Bose-condense. Instead,
in the models discussed here, because of the enhanced pair
mobility due to the undressing process, the transition to the superconducting
state occurs through pair binding rather than Bose-condensation,
except in extremely low carrier concentration regimes\cite{bose}.
Finally, we believe that the relevant boson excitations that were
described here through the generalized Holstein model are likely
to be electronic excitations of the atomic 
shells\cite{hole,holesc1,diat,diat2} rather than
related to ionic lattice displacements as in Alexandrov's work.

In summary, we believe that the basic physics discussed here is relevant to
the understanding of high $T_c$ superconductivity in cuprates.
Because of its generality it should also
apply to any Fermi liquid system where the quasiparticle dressing
decreases as the local carrier concentration increases, and we suggest that 
this may be generally the case for metals for which  at least some of the
quasiparticles in the normal state have hole-like character.
This is also consistent with the empirical observation that superconductivity
appears to be quite generally associated with positive values of the Hall
coefficient in the normal state at least in some 
directions.\cite{chapnik,correl}.
Thus, the possibility that this physics
may play an essential role in $all$ superconductors should not
be excluded\cite{other}. Note, however, that the physics of undressing
discussed here leads uniquely to the symmetry of the superconducting state
being s-wave\cite{holesc2}.

\acknowledgements
The author is grateful to D. Basov, M. Norman and H. Suhl for stimulating
discussions.

\begin{figure}
\caption { 
Phenomenology of high $T_c$ superconductors (schematic) as
described by the models in this paper.
Heavily dressed quasiparticles at low concentrations undress as the
temperature is lowered or as the carrier concentration increases.
$s.t.$ denotes spectral weight transfer (from high to low
frequencies in the direction of the arrows), or $strip$ $tease$. Below the curve
labeled $T_c$ the system is superconducting. The initial rise
in $T_c$ versus $n$ is due to the increasing number of carriers.
}
\label{Fig. 1}
\end{figure}

\begin{figure}
\caption { The physics of electron-hole asymmetric polarons. A boson degree
of freedom is associated with each site. The first electron at the site
causes a small change in the ground state of this degree of freedom, the 
second electron causes a large change. For holes, the situation is
reversed. Two examples of the boson degree of freedom are shown, an 
oscillator and a spin 1/2.
}
\label{Fig. 2}
\end{figure}

\begin{figure}
\caption { Schematic depiction of hopping processes for different
band fillings. As the (electron) band filling increases, the
hopping amplitude decreases and carriers become heavier.
$n_h$ denotes hole concentration.
}
\label{Fig. 3}
\end{figure}

\begin{figure}
\caption { Electronic band versus doping (schematic) as predicted
by Eq. (13). As the hole doping increases, the Fermi level
moves down and the band becomes broader.
}
\label{Fig. 4}
\end{figure}

\begin{figure}
\caption { Spectral function for a site for the generalized
Holstein model. The lowest peak gives the quasiparticle weight,
the other peaks give the incoherent contribution which in an
extended system will broaden into a smooth spectrum, given by the envelope 
dashed lines. (a) corresponds to Eq. (44a), with $n_\downarrow=0$,
and (b) to Eq. (44b) with $n_\downarrow=1$. As the hole concentration increases 
the weight of the
quasiparticle (q.p.) peak increases and the incoherent contribution
shifts to lower frequencies and decreases in total weight.
Parameters used are $g^2=3$, $\gamma=0.5$.
}
\label{Fig. 5}
\end{figure}

\begin{figure}
\caption { 
Angle resolved photoemission spectrum for low (solid lines) and high
(dashed lines) hole concentrations, from Eq. (61). For each case two
values of the momentum are shown, $\epsilon_k-\mu=10meV$ and
 $\epsilon_k-\mu=20meV$. Parameters used are $g^2=4$, $\gamma=0.75$, 
$\omega_0=10meV$, $\sigma_\omega=10meV$.
}
\label{Fig. 6}
\end{figure}


\begin{references}
\bibitem{ding} H. Ding et al, cond-mat/0006143 and references therein.
\bibitem{coh1} T. Takahashi et al, Physica B {\bf 165\&166}, 1221 (1990).
\bibitem{coh2} D.S. Dessau et al, Phys. Rev. Lett. {\bf 66}, 2160 (1991). 
\bibitem{coh3} A.V. Puchkov et al,  Phys. Rev. B {\bf 54}, 6686 (1996).
\bibitem{color} J. E. Hirsch, Physica C {\bf 201}, 347 (1992). 
\bibitem{polar1} J. E. Hirsch, in "Polarons and Bipolarons in high-$T_c$ Superconductors and 
Related Materials", ed. by E.K.H. Salje, A.S. Alexandrov and W.Y. Liang,
Cambridge University Press, Cambridge, 1995, p. 234 .
\bibitem{area} J. E. Hirsch, Physica C {\bf 199}, 305 (1992);
 J.E. Hirsch and F. Marsiglio, Physica C {\bf 331}, 150 (2000);
cond-mat/0004496, submitted to Phys.Rev. B.
\bibitem{basov} D.N. Basov et al, Science {\bf 283}, 49 (1999);
A.S. Katz et al, Phys.Rev. B {\bf 61}, 5930 (2000); D.N. Basov et al, preprint.
\bibitem{fugol}I. Fugol et al, Sol.St.Comm. {\bf 86}, 385 (1993).
\bibitem{hole} J. E. Hirsch, Phys.Lett. A{\bf 134}, 451 (1989). 
\bibitem{holesc1} J.E. Hirsch and S. Tang, Sol.St.Comm. {\bf 69}, 987 (1989);
Phys. Rev. B {\bf 40}, 2179 (1989);
 J.E. Hirsch and F. Marsiglio, Phys. Rev. B {\bf 41}, 2049 (1990).
\bibitem{holesc2} J.E. Hirsch and F. Marsiglio, Phys. Rev. B {\bf 39}, 11515
(1989); Phys. Rev. B {\bf 45}, 4807 (1992); Physica C {\bf 162-164}, 591 (1989);
F. Marsiglio and J.E. Hirsch, Phys. Rev. B {\bf 41}, 6435 (1990).
\bibitem{tun} F. Marsiglio and J.E. Hirsch, Physica C {\bf 159}, 157 (1989);
J.E. Hirsch, Phys. Rev. B {\bf 59}, 11962 (1999).
\bibitem{imb} J.E. Hirsch, Phys. Rev. Lett. {\bf 72}, 558 (1994);
Phys. Rev. B {\bf 58}, 8727 (1998).
\bibitem{twoorb} J.E. Hirsch, Phys. Rev. B {\bf 43}, 11400 (1991).
\bibitem{holst} T. Holstein, Ann. Phys. (N.Y.)  {\bf 8}, 325 (1959).
\bibitem{pincus} P. Pincus, Sol. St. Comm.{\bf 11}, 51 (1972).
\bibitem{polar2} J.E. Hirsch, Phys. Rev. B {\bf 47}, 5351 (1993).
\bibitem{norman} M.R. Norman, M. Randeria, B. Janko and
J.C. Campuzano, Phys. Rev. B {\bf 61}, 14742 (2000).
\bibitem{alex1} A.S. Alexandrov, Phys. Rev. B {\bf 61}, 12315 (2000)
and references therein.
\bibitem{alex2} A.S. Alexandrov and J. Ranninger, Physica C {\bf 198}, 360 (1992).
\bibitem{uchida} S. Uchida et al, Phys. Rev. B {\bf 43}, 7942 (1991).
\bibitem{diat} J.E. Hirsch, Phys. Rev. B {\bf 48}, 3340 (1993).
\bibitem{sham} L.J. Sham, Sol.St. Comm. {\bf 20}, 623 (1976).
\bibitem{lang} I.G. Lang and Y.A. Firsov, Sov.Phys. JETP {\bf 16}, 1301 (1963).
\bibitem{mahan} G.D. Mahan, "Many-Particle Physics", Plenum, New York, 1981.
\bibitem{dilute} F. Marsiglio and J. E. Hirsch, Physica C {\bf 171}, 554 (1990). 
\bibitem{strong} J. E. Hirsch, Physica C {\bf 161}, 185 (1989).  
\bibitem{strongweak} F. Marsiglio and J. E. Hirsch, Physica C {\bf 165}, 71 (1990).  
\bibitem{ander}P.W. Anderson, Science {\bf 268}, 1154 (1995);
S. Chakravarty, H.Y. Young and E. Abrahams, Phys.Rev.Lett. {\bf 82}, 2366 (1999).
\bibitem{kivel}V.J. Emery , S.A. Kivelson and J.M. Tranquada,
cond-mat/9907228, to appear in Proc. Natl. Acad. Sci. USA .
\bibitem{bose} J. E. Hirsch, Physica C {\bf 179}, 317 (1991).
\bibitem{diat2} J. E. Hirsch, Chem.Phys.Lett. {\bf 171}, 161 (1990).
\bibitem{chapnik}I.M. Chapnik, Phys.Lett. {\bf 72A}, 255 (1979).
\bibitem{correl} J.E. Hirsch, Phys. Rev. B {\bf 55}, 9007 (1997).
\bibitem{other} J. E. Hirsch, Physica C {\bf 158}, 326 (1989);
Phys.Lett. A{\bf 138}, 83 (1989).


\end{references}
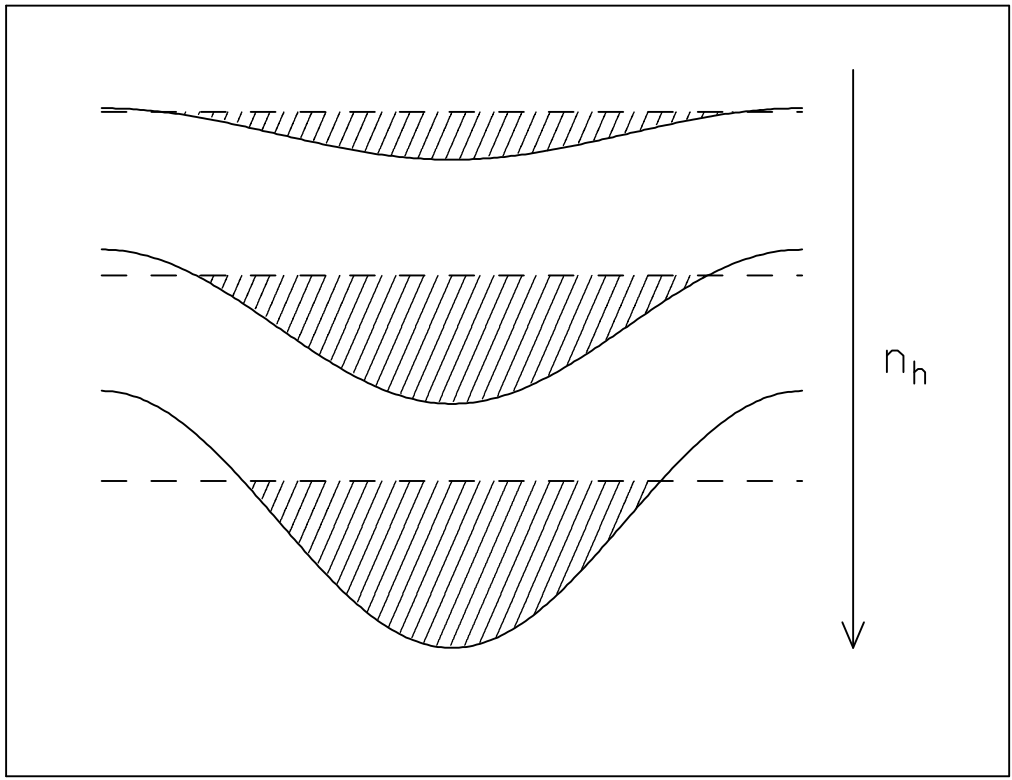
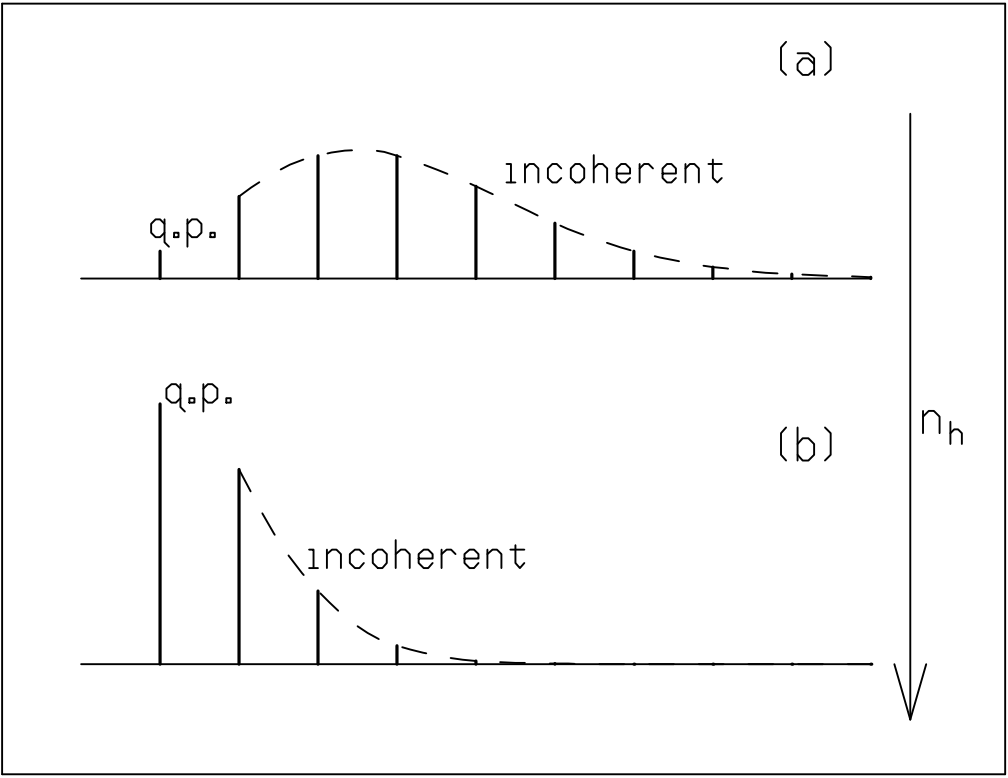
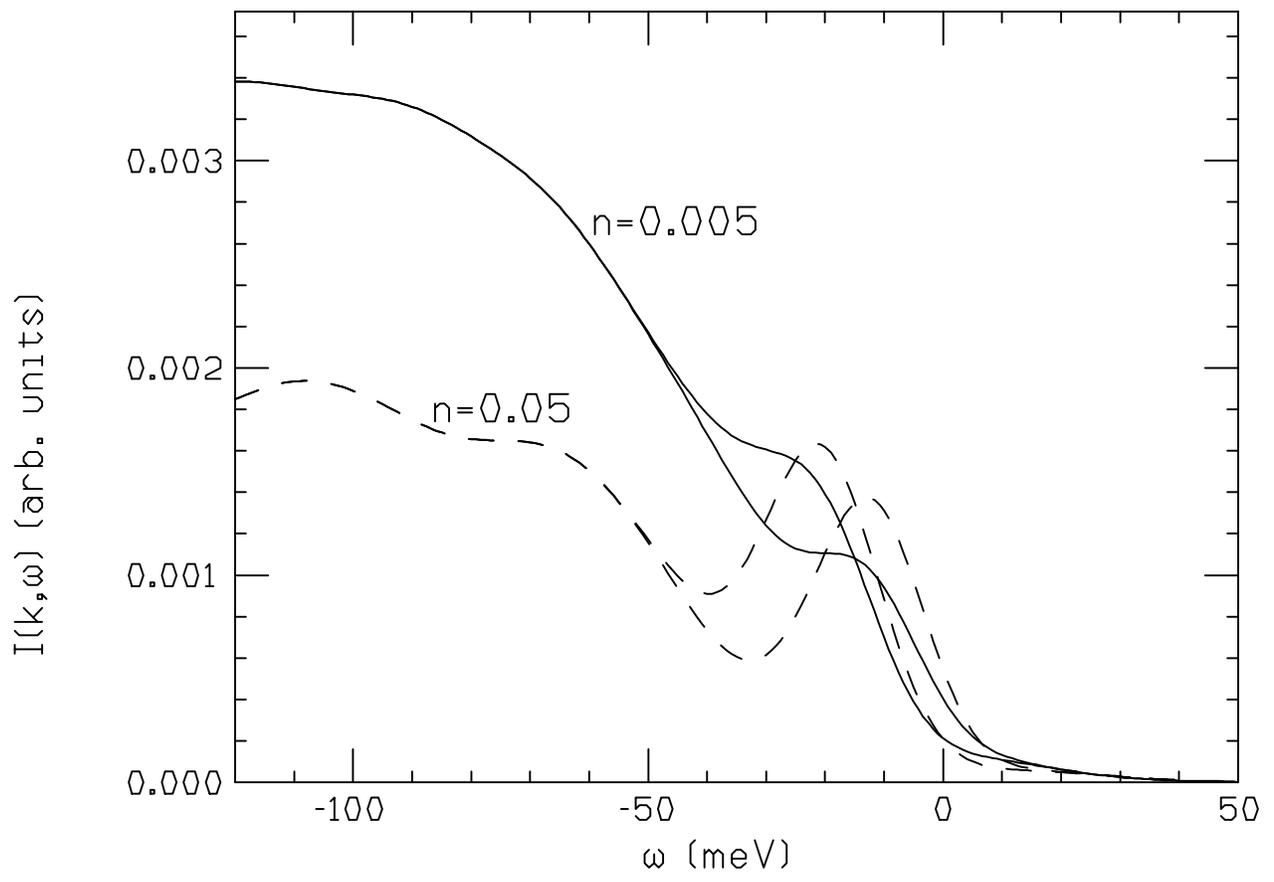
\end{document}